%% file: root.tex
\documentclass[letterpaper, 10 pt, conference]{ieeeconf}  

\IEEEoverridecommandlockouts                              

\usepackage{amsmath} 
\usepackage{amssymb}  
\usepackage{relsize}
\usepackage{bm}
\usepackage{float}
\usepackage{algorithm}
\usepackage{algpseudocode}
\usepackage{dutchcal}
\usepackage{cuted}
\usepackage{siunitx}
\usepackage[short]{optidef}
\usepackage{xcolor}
\usepackage{booktabs}
\usepackage{multirow}
\usepackage{colortbl}
\usepackage{array}
\usepackage{url}
\usepackage{svg}
\usepackage[11pt]{moresize}

\newcolumntype{H}{>{\setbox0=\hbox\bgroup}c<{\egroup}@{}} 

\usepackage[mode=buildnew]{standalone}

\title{\LARGE \bf
A Stochastic Nonlinear Model Predictive Control \\with an Uncertainty Propagation Horizon \\for Autonomous Vehicle Motion Control}

\author{Baha Zarrouki$^{1,2}$, Chenyang Wang$^{2}$ and Johannes Betz$^{2}$
\thanks{$^{1}$ Chair of Automotive Technology, Technical University Munich}
\thanks{$^{2}$ Professorship of Autonomous Vehicle Systems, TUM School of Engineering and Design, Technical University Munich, 85748 Garching, Germany; Munich Institute of Robotics and Machine Intelligence (MIRMI), \{{baha.zarrouki}, {16chenyang.wang}, {johannes.betz}\}@tum.de
}}%

\begin{document}
\maketitle
\thispagestyle{empty}
\pagestyle{empty}

\begin{abstract}
Employing Stochastic Nonlinear Model Predictive Control (SNMPC) for real-time applications is challenging due to the complex task of propagating uncertainties through nonlinear systems. This difficulty becomes more pronounced in high-dimensional systems with extended prediction horizons, such as autonomous vehicles. 
To enhance closed-loop performance in and feasibility in SNMPCs, we introduce the concept of
the Uncertainty Propagation Horizon (UPH). The UPH limits the time for uncertainty propagation through system dynamics, preventing trajectory divergence, optimizing feedback loop advantages, and reducing computational overhead. Our SNMPC approach utilizes Polynomial Chaos Expansion (PCE) to propagate uncertainties and incorporates nonlinear hard constraints on state expectations and nonlinear probabilistic constraints. We transform the probabilistic constraints into deterministic constraints by estimating the nonlinear constraints' expectation and variance.
We then showcase our algorithm's effectiveness in real-time control of a high-dimensional, highly nonlinear system—the trajectory following of an autonomous passenger vehicle, modeled with a dynamic nonlinear single-track model.
Experimental results demonstrate our approach's robust capability to follow an optimal racetrack trajectory at speeds of up to $37.5\unit{\meter\per\second}$ while dealing with state estimation disturbances, achieving a minimum solving frequency of $97 \unit{\hertz}$. Additionally, our experiments illustrate that limiting the UPH renders previously infeasible SNMPC problems feasible, even when incorrect uncertainty assumptions or strong disturbances are present.
\end{abstract}
\section{Introduction}
\input{sections/Intro}
\section{Nominal NMPC}
 \label{sec:nominal_nmpc}
\input{sections/nominal_nmpc}

\section{Stochastic NMPC}
 \label{sec:stochastic_nmpc}
\input{sections/stochastic_nmpc}

\section{Stochastic Nonlinear Model Predictive Control for Trajectory Following of Autonomous Vehicles}
\label{sec:traj_following}
\input{sections/traj_following}

\section{Simulation Results}
\label{sec:sim_results}
\input{sections/sim_results}

\section{Conclusions and Future Work}
\label{sec:conclusion}
To address the challenge of infeasibility arising from the unrestricted propagation of uncertainties through system dynamics in real-time SNMPCs, we introduce the concept of the Uncertainty Propagation Horizon (UPH) combined with Polynomial Chaos Expansion. A short UPH strategically restricts the propagation of uncertainty samples, preventing divergence in trajectories, leveraging the benefits of the feedback loop, and minimizing unnecessary computations. \\
In the context of following a raceline trajectory with speeds up to $37.5\unit{\meter\per\second}$, we demonstrate the robustness of SNMPC by showcasing its ability to maintain desired performance despite emulated state estimation disturbances. \\This robustness is evident in the slight degradation of only $1.6\%$ compared to nominal SNMPC conditions and an impressive $61\%$ improvement over nominal NMPC in terms of maximum lateral deviation. These results underscore SNMPC's potential for reliable vehicle guidance in real-world scenarios and its critical role in preventing extreme deviations, thereby enhancing the safety of autonomous vehicles.
Furthermore, our experimental findings reveal that limiting the UPH transforms infeasible standard SNMPC problems into feasible ones, even when faced with incorrect uncertainty assumptions or strong disturbances.
Finally, our approach makes real-time optimal control problems for nonlinear systems computationally feasible, enabling the rapid solution of stochastic nonlinear optimal control problems at a minimum frequency of $97 \unit{\hertz}$.\\
Our approach generally outperforms the nominal NMPC, especially as disturbances increase in magnitude, but shows only a slight advantage in cases of very small disturbances. Also, the SNMPC approach remains sensitive to the assumed uncertainty standard deviations and the length of the UPH. In the future, adaptive methods can continuously assess and refine uncertainty assumptions using real-time sensor data and system behavior. The ideal UPH length can vary in different operating conditions and disturbance scenarios. By implementing adaptability, SNMPC can autonomously select the most appropriate UPH length for specific situations. 
\bibliographystyle{IEEEtran}
\bibliography{literatur} 

\end{document}

%% file: sections/Intro.tex









Model uncertainties and external disturbances can substantially impact Model Predictive Control (MPC) algorithms, leading to suboptimal solutions and poor closed-loop performance. To handle these challenges, the MPC problem can be designed to ensure the satisfaction of the state and control constraints for all possible disturbance sequences, known as robust MPC, or with a prespecified probability, termed stochastic MPC \cite{rawlings2017model}. The challenge becomes even more pronounced when dealing with nonlinear systems.
Robust Nonlinear MPC (NMPC), such as tube-based NMPC \cite{bemporad2007robust} or robust min-max NMPC \cite{chen1997game}, ensures system stability and performance under worst-case perturbations. However, this worst-case scenario approach can be overly conservative when the likelihood of such events is low, potentially resulting in suboptimal closed-loop performance. Additionally, robust NMPC may fail to guarantee constraint satisfaction when disturbances occur outside the assumed range. In contrast, stochastic NMPC (SNMPC) leverages probabilistic uncertainty descriptions, allowing the incorporation of acceptable risk levels in system operation. This approach strikes a balance between closed-loop performance and constraint violations, mitigating the inherent conservativeness associated with robust NMPC \cite{mesbah2016stochastic}.\\
SNMPC has faced developmental challenges, mainly due to the intricate task of efficiently propagating uncertainty through complex nonlinear systems. Various methods have been explored to address this issue. The Gaussian-mixture approximation method \cite{weissel2009stochastic} characterizes transition probability distributions of states. However, it is less suitable for handling time-invariant uncertainties, such as system parameters and initial conditions. An alternative approach for efficient uncertainty propagation is Polynomial Chaos Expansion (PCE) \cite{fagiano2012nonlinear} \cite{mesbah2014stochastic}. PCE deals with time-invariant uncertainties, offering efficient means to predict the temporal evolution of probability distribution moments for stochastic states. Furthermore, the work presented in \cite{paulson2019efficient} extends this framework to accommodate time-variant uncertainties. 
Notably, \cite{bradford2018stochastic} and \cite{bradford2020stochastic} introduced an alternative propagation framework employing Gaussian Regression. To capture better the distribution of stochastic states, work \cite{bradford2021combining} further combined Gaussian Regression and PCE, which is however computationally intensive and impractical for real-time systems. 
Our work is inspired by \cite{fagiano2012nonlinear}, yet it extends and enhances several critical aspects of the approach. While \cite{fagiano2012nonlinear} relies on control correction as a decision variable, our approach directly computes control inputs within the optimal control problem, eliminating the need for pre-computed state-feedback control laws. Additionally, Approach \cite{fagiano2012nonlinear} does not explicitly consider system dynamics in the optimal control problem, relying on nominal states for the pre-computed control law. In contrast, we employ PCE-estimated states, avoiding the need for nominal states' evolution. We also introduce probability constraints and transform them into deterministic counterparts and extend the optimization problem with a terminal cost. 
Furthermore, this work also draws inspirations from work \cite{mesbah2014stochastic} and extends its concepts.  In contrast to the original work, which focused on formulating linear chance inequality constraints applied solely to state expectations, we now account for nonlinear chance constraints by estimating the nonlinear constraints' expectation and variance with PCE.
Instead of evolving nominal states over time while imposing hard nonlinear inequality constraints on them, we evolve only samples around the system's initial state variables, incorporating uncertainty through the computed expectation using PCE, i.e. we discard the nominal state after the initial time step, considering it only for sampling. Additionally, we introduce hard linear and nonlinear constraints on the states' expectations, effectively accounting for the uncertainties.\\
Prior to our work, SNMPC remained unexplored in motion control applications.
In real-time context, linear SMPCs are the standard; for instance, \cite{knaup2023safe} controls linearized vehicle systems using covariance steering. The challenge with SNMPC in such applications lies in the infeasibility arising from rapidly expanding variances in states and constraints. Approaches like \cite{BRADFORD2018417} addressed this using unscented transformation and a specialized Robust Horizon but may not resolve infeasibility when SNMPC employs alternative methods like PCE. In this work, we propose a novel and more general concept —the Uncertainty Propagation Horizon (UPH) that helps avert trajectory divergence, leverages the advantages of the feedback loop, and reduces unnecessary computations while estimating both states and nonlinear constraints.
In summary, this work presents three main contributions:
\begin{enumerate}
    \item We formulate a general PCE-SNMPC incorporating state expectations' in the cost function and in the linear hard constraints. It handles both nonlinear hard constraints on state expectations and estimated nonlinear probabilistic constraints in expectation and variance. 
        \item We introduce the novel UPH concept and demonstrate its effectiveness in addressing infeasibility issues, even in the presence of incorrect uncertainty assumptions or strong disturbances.
        \item We showcase our SNMPC's robustness and real-time capabilities by applying it to control a high-dimensional and highly nonlinear system affected by severe disturbances —a full-scale autonomous vehicle.
\end{enumerate}

%% file: sections/nominal_nmpc.tex
We employ the following notation: Given a variable $z \in \mathbb{R}$, we define $\boldsymbol{z} = [z_0, z_1, \ldots, z_n]^T \in \mathbb{R}^n$ as the vector composed of $z$ variables. Additionally, $\boldsymbol{Z} = [\boldsymbol{z}^{(0)}, \boldsymbol{z}^{(1)}, \ldots, \boldsymbol{z}^{(m)}] \in \mathbb{R}^{n \times m}$ represents the matrix consisting of concatenated vectors $\boldsymbol{z}$.

In this work, we consider the following nominal NMPC: 
\begin{equation}
\begin{aligned}
& \textbf{Problem 1} && \textbf{Nominal NMPC}\\ 
&  \underset{\boldsymbol{x}(.), \boldsymbol{u}(.)}{\min} & & 
\begin{aligned}
    \int^{T_p}_{\tau=0} & l(\boldsymbol{x}(\tau),\boldsymbol{u}(\tau)) \space  d\tau \\ 
 & + m(\boldsymbol{x}(T_p))
\end{aligned}
  \\ 
& \text{subject to} & & \boldsymbol{x}_{0} \leq \boldsymbol{x}(0) \leq \boldsymbol{x}_{0} \text {, }  \\
& & & \dot{\boldsymbol{x}}(t) = f(\boldsymbol{x}(t),\boldsymbol{u}(t)) \text {, } & t \in[0, T_p), \\
& & & \underline{\boldsymbol{h}} \leq h(\boldsymbol{x}(t), \boldsymbol{u}(t)) \leq \bar{\boldsymbol{h}},& t \in[0, T_p), \\
& & & \underline{\boldsymbol{x}} \leq J_{\mathrm{bx}} \space \boldsymbol{x}(t) \leq \bar{\boldsymbol{x}}, & t \in[0, T_p), \\
& & & \underline{\boldsymbol{u}} \leq J_{\mathrm{bu}} \space \boldsymbol{u}(t)\leq \bar{\boldsymbol{u}}, & t \in[0, T_p), \\
& & & \underline{\boldsymbol{h}}^{\mathrm{e}} \leq h^{\mathrm{e}}(\boldsymbol{x}(T_p)) \leq \bar{\boldsymbol{h}}^{\mathrm{e}}, \\
& & & \underline{\boldsymbol{x}}^{\mathrm{e}} \leq J_{\mathrm{bx}}^{\mathrm{e}} \space \boldsymbol{x}(T_p) \leq \bar{\boldsymbol{x}}^{\mathrm{e}}, \\
\end{aligned}
\label{eq:nominal NMPC problem}
\end{equation}
Here, $\boldsymbol{x} \in \mathbb{R}^{n_x}$ denotes the state vector, $\boldsymbol{u} \in \mathbb{R}^{n_u}$ the control vector, $t$ the discrete time, $T_p$ the prediction horizon, $f$ the system dynamics, $h$ and $h^e$ the path and terminal nonlinear inequality constraints, $J_{bx}$ and $J_{bx}^e$ help express linear path and terminal state constraints, $J_{bu}$ helps express control input constraints and $x_0$ the initial state.
Also, $l: \mathbb{R}^{n_{\mathrm{x}}} \times \mathbb{R}^{n_{\mathrm{u}}}  \rightarrow \mathbb{R}$ denotes the stage cost and $m: \mathbb{R}^{n_{\mathrm{x}}}  \rightarrow \mathbb{R}$ the terminal cost.


%% file: sections/stochastic_nmpc.tex
In this work, the SNMPC effectively manages uncertain nonlinear system dynamics by assuming that the true system state follows a known distribution around the measured state. 
To achieve this, we sample states in the vicinity of the measured state and simulate their dynamics behavior at each shooting node while solving the MPC problem. 
These sampled states, characterized by uncertainty, allow the MPC to consider estimated state variables and effectively satisfy linear- and nonlinear constraints during its optimization. 
However, propagating uncertainties across the entire prediction horizon can render the optimization infeasible for complex nonlinear systems. To mitigate this issue, we conceive an SNMPC that limits the propagation of uncertain state samples and constraints to a defined horizon, which we refer to as the Uncertainty Propagation Horizon (UPH).\\
In this section, we briefly explain the Polynomial Chaos Expansion (PCE) method for approximating uncertain variables. Next, we introduce the concept of the UPH and describe how it governs the propagation of sampled states and constraints. Additionally, we demonstrate how the SNMPC with probabilistic constraints can be transformed into a deterministic problem. Finally, we present our SNMPC problem formulation and algorithm for online application. 
\subsection{Polynomial Chaos Expansion}
PCEs are able to approximate any stochastic variable $x_w$ using second-order moments \cite{wiener1938homogeneous} using the $L_2$-convergent expansion. This expansion relies on a series of multivariate orthogonal polynomials $\Phi_{\alpha_k}$ \cite{cameron1947orthogonal}, each associated with a vector of indices $\boldsymbol{\alpha}_k = [\alpha_{1,k}, \alpha_{2,k}, \ldots, \alpha_{n,k}]$:
\begin{equation}
    x_w(\boldsymbol{w}) = \sum_{k=0}^{\infty} c_k \Phi_{\boldsymbol{\alpha}_k}(\boldsymbol{w})
    \label{eq:x_w_full}
\end{equation}
Here, $\boldsymbol{w}$ represents uncertain system parameters, and $c_k$ denotes the corresponding PCE coefficients. Each multivariate polynomial in this expansion is a product of $n$ univariate polynomials, each having a degree denoted as $\alpha_{i,k}$, where $i \in {1, 2, \ldots, n}$. The sum of polynomial degrees across all dimensions is represented as $d_{PC,tot}$.
In practice, the polynomial series is truncated to a maximal total degree $d_{PC,max}$. The total number of terms, denoted as $L$, in this truncated PC expansion can be computed based on the maximum total expansion degree $d_{PC,max}$, where $d_{PC,tot} \leq d_{PC,max}$, and the total number of uncertain system parameters $n_w$:
\begin{equation}
    L = \frac{(n_w + d_{PC,max})!}{n_w! d_{PC,max}!}
\end{equation}
As a result, the truncated PCE series from Equation (\ref{eq:x_w_full}) simplifies to:
\begin{equation}
    \hat{x}_w(\boldsymbol{w}) = \sum_{k=0}^{L-1} c_k \Phi_{\boldsymbol{\alpha}_k}(\boldsymbol{w}) = \boldsymbol{\Phi}(\boldsymbol{w}) \boldsymbol{c}
    \label{eq:truncated_PCE}
\end{equation}
The choice of the polynomial basis, such as Hermite, Legendre, or others, depends on the assumed distribution of the random variable, such as Gaussian, uniform, etc \cite{fagiano2012nonlinear}.
\subsection{Uncertainty Propagation Horizon (UPH)}
Simulating uncertainty propagation throughout the entire prediction horizon in the PCE-based SNMPC can often result in overly conservative solutions and even to divergence in the trajectories of propagated uncertain states. This approach may lead to infeasible problem instances or sub-optimal solutions, ultimately diminishing the closed-loop system's performance. It also involves unnecessary computations for estimating uncertainties far into the future, which underutilizes the benefits of the feedback loop. \\
To address this issue, we introduce the novel Uncertainty Propagation Horizon (UPH) concept. The UPH represents the period within the prediction horizon, i.e., $T_u \leq T_p$, over which uncertainties are explicitly considered and propagated within the SNMPC framework. Upon reaching the Uncertainty Propagation Horizon (UPH), only the last estimated variables at $t = T_u$ are propagated until $T_p$.\\
\subsection{Uncertain States Propagation}
\label{subsec:uncertain_states_prop}
\begin{figure}[ht]
\centering
\includegraphics[width=.48\textwidth]{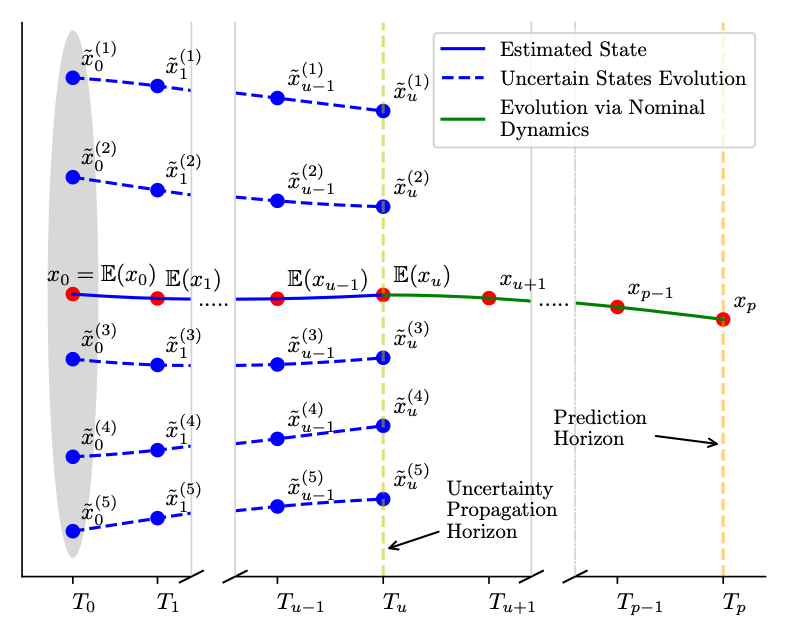}
\caption{Evolution of states within the propagation- and prediction horizon.}
\label{fig:propagation_horizon}
\end{figure}

Figure \ref{fig:propagation_horizon} depicts how the collocation points $\tilde{x}_0^{(i)}$ are sampled from the distribution of initial state and evolved through system dynamics from $t=T_0$ to $t = T_u$, here under the same control inputs. At each time step before the UPH, the expectation of states is estimated by PCE. After UPH, the evolution of collocation points stops, and the latest computed expectation $\mathbb{E}(x_u)$ evolves via nominal dynamics until prediction horizon $t=T_p$.

We construct the vector $\boldsymbol{\Tilde{X}} \in  \mathbb{R}^{n_s} \times \mathbb{R} ^{n_x}$ containing $n_s$ generated state samples such as $\boldsymbol{\Tilde{X}} = [\boldsymbol{\Tilde{x}}^{(1)}, \ldots, \boldsymbol{\Tilde{x}}^{(n_s)}]^T$. The evolution of these samples is governed by the equation:
\begin{equation}
    \boldsymbol{\Tilde{X}}_{t+1} = 
    \begin{cases}
    f(\boldsymbol{\Tilde{X}}_{t},\boldsymbol{u}) , & \text{if } t\! \in \! 
			\left\{0,...,N_{u-1}\right\}  \\
    \boldsymbol{O}_{n_s \times n_x},  & \text{if } t\! \in \! 
			\left\{N_u,...,N_{p-1}\right\}  \\ 
\end{cases}
\end{equation}
Here, $N_u = \frac{T_u}{T_s}$ and $N_p = \frac{T_p}{T_s}$ represent the lengths of the uncertainty propagation and prediction horizons, respectively, with $T_s$ being the MPC's sampling time.
We compute then the system states' expectations for the SNMPC problem as follows:
\begin{equation}
     \mathbb{E}[\boldsymbol{x}_{t}]= 
    \begin{cases}
     \boldsymbol{c}^{(\boldsymbol{x})}_0, & \text{if } t\! \in \! 
			\left\{0,...,N_{u-1}\right\}  \\
    \boldsymbol{x}_{t} = f(\mathbb{E}[\boldsymbol{x}_{t-1}],\boldsymbol{u}),  & \text{if } t\! \in \! 
			\left\{N_{u},...,N_{p-1}\right\}  \\
\end{cases}
\label{eq:x dynamics cases}
\end{equation}
Here, $\boldsymbol{C}^{(x)}= [\boldsymbol{c}^{(\boldsymbol{x})}_0, \ldots, \boldsymbol{c}^{(\boldsymbol{x})}_{L-1}]$ is the vector containing the PCE coefficients and $\boldsymbol{c}^{(\boldsymbol{x})}_i \in \mathbb{R}^{n_x}, \forall i \in \{0, \ldots, L-1\}$ such that:
\begin{equation}
    \boldsymbol{c}^{(\boldsymbol{x})}_i = [c^{(x_1)}_i, \ldots, c^{(x_{n_x})}_i]^T
\end{equation}
In other words, $\boldsymbol{c}^{(\boldsymbol{x})}_0$ contains the coefficients of the 0th order polynomial in the PCE. Similar to the Monte Carlo methods, we estimate the expectation in (\ref{eq:x dynamics cases}) by running a finite number of simulations. 
The PCE coefficients are computed using the matrix $A \in \mathbb{R}^L \times \mathbb{R}^{n_s}$ with the following relationship:
\begin{equation}
\begin{aligned}
     \boldsymbol{C}^{(\boldsymbol{x})} = A \cdot\boldsymbol{\Tilde{X}}_{t+1}\\
    A = (\boldsymbol{\tilde{\Phi}}^T\boldsymbol{\tilde{\Phi}})^{-1}\boldsymbol{\tilde{\Phi}}^T 
\end{aligned}
\label{eq:PCE_x coefficients}
\end{equation}
Equation \ref{eq:PCE_x coefficients} is the solution of an $l_2$-norm least-squares regression problem:
\begin{equation}
    \begin{aligned}
&  \underset{\boldsymbol{C}^{(\boldsymbol{x})}}{\min} & &  ||\boldsymbol{\Tilde{X}}-\boldsymbol{\tilde{\Phi}}\boldsymbol{C}^{(\boldsymbol{x})}||_2^2
\end{aligned}
\end{equation}
The initial samples matrix $\boldsymbol{\Tilde{X}}_{0}$ is constructed as:
\begin{equation}
     \boldsymbol{\Tilde{x}}^{(i)} = \boldsymbol{x_0} + \boldsymbol{\tilde{w}}^{(i)}, i \in \{1, \ldots, n_s\}  
\end{equation}
Where $\boldsymbol{\tilde{W}} = [\boldsymbol{\tilde{w}}^{(1)}, \ldots, \boldsymbol{\tilde{w}}^{(n_s)}]$ contains $n_s$ sampled disturbances, e.g., using Hammersley low-discrepancy sequence, taking into account the assumed standard deviations $\boldsymbol{\sigma}_w$.
The matrix $\boldsymbol{\tilde{\Phi}} \in \mathbb{R}^{n_s} \times \mathbb{R}^L $ is defined as:
\begin{equation}
    \boldsymbol{\tilde{\Phi}} = [\boldsymbol{\Phi}(\boldsymbol{\Tilde{w}}^{(1)}), \ldots , \boldsymbol{\Phi}(\boldsymbol{\Tilde{w}}^{(n_s)})]^T
\end{equation}
The vectors $ \boldsymbol{\Phi}(\boldsymbol{\Tilde{w}}^{(i)})$ are the multivariate polynomials evaluated at $\boldsymbol{\Tilde{w}}^{(i)}$. Each vector is defined as:
\begin{equation}
    \boldsymbol{\Phi}(\boldsymbol{\Tilde{w}}^{(i)}) :=  [\Phi_{\boldsymbol{\alpha}_0}(\boldsymbol{\Tilde{w}}^{(i)}), \ldots, \Phi_{\boldsymbol{\alpha}_{L-1}}(\boldsymbol{\Tilde{w}}^{(i)})] , i \in \{1, \ldots, n_s\}
\end{equation}
The multivariate polynomials $\boldsymbol{\Phi}_{\boldsymbol{\alpha}_k}$ are computed using chosen polynomials, e.g., Hermite for an assumed Gaussian distribution \cite{olver2010nist}, and based on the polynomial degrees $\alpha_k$ (\ref{eq:truncated_PCE}). These degrees are determined by the number of variables affected by uncertainty and the expansion degree $d_{PC,max}$. 
It's important to note that the polynomial degrees are fixed and computed offline. However, the sample points are updated at each MPC step, as we assume that disturbance assumptions and, consequently, the sample points can change over time when the system operates online. We plan to adapt the disturbance assumptions online and account for real-time sensor feedback in future work.


\subsection{Uncertainty and Nonlinear Chance Constraints}
\label{subsec:uncertain_nonlinear_ineq_prop}
\begin{figure}[ht]
\centering
\includegraphics[width=.48\textwidth]{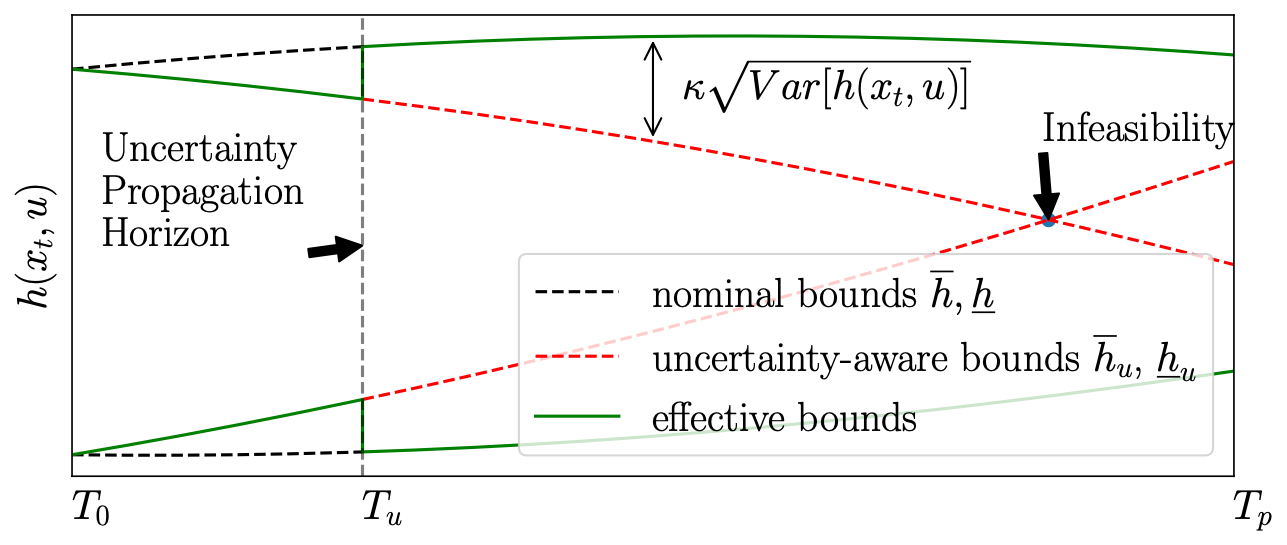}
\caption{General concept of nonlinear inequality constraint's evolution.}
\label{fig:constraint_infeasibility}
\end{figure}

Figure \ref{fig:constraint_infeasibility} provides a visual representation of how the bounds of nonlinear inequality constraints change over time.
The propagation of uncertainty leads to increasing variances in the states, which can eventually result in infeasibility. 
However, the UPH helps restore the effective bounds of inequality constraints to their nominal values after UPH, thus enhancing problem feasibility.
In the context of stochastic MPC, the nonlinear constraints in Problem 1 are reformulated as chance constraints, expressed as:
\begin{equation}
    \mathbb{P}(\underline{\boldsymbol{h}} \leq h(\boldsymbol{x}(t), \boldsymbol{u}(t)) \leq \bar{\boldsymbol{h}}) \geq 1 - p
\end{equation}
Here, $p \in (0,1]$ denotes the desired probability of constraint violation w.r.t. $h$. To improve computational efficiency in solving the control problem\cite{mesbah2014stochastic}, the nonlinear probability constraints inequality are transformed into deterministic constraints \cite{bradford2020stochastic}:
\begin{equation}
\begin{split}
    h_u(\boldsymbol{x},\boldsymbol{u}) &= \mathbb{E}[h(\boldsymbol{x},\boldsymbol{u})] + \kappa\sqrt{\text{Var}[h(\boldsymbol{x},\boldsymbol{u})]}
\end{split}
\label{eq:deterministic h}
\end{equation}
Where  $\kappa = \sqrt{(1- p)/p}$. The first- and second-order moments of $h$ are given by:
\begin{equation}
\mathbb{E}[h(\boldsymbol{x},\boldsymbol{u})] = 
\begin{cases}
     c^{(h)}_0, &\text{if } t\! \in \! 
			\left\{0,...,N_{u-1}\right\}  \\
     h(\mathbb{E}[\boldsymbol{x}_t],\boldsymbol{u}),  &\text{if } t\! \in \! 
			\left\{N_{u},...,N_{p-1}\right\}  \\
\end{cases}
\label{eq:E(h) dynamics cases}
\end{equation} 
\begin{equation}
 \text{Var}[h(\boldsymbol{x},\!\boldsymbol{u})]\! = \!
\begin{cases}
\begin{aligned}
    &\sum_{k=1}^{L-1}\!
    (c^{(\!h\!)}_k)^2 \mathbb{E}[\Phi\!_{\alpha\!_k}\!(\!\boldsymbol{w}\!)^2], 
    &&\text{if } t\! \in \! 
			\left\{0,...,N_{u-1}\right\}  \\
    &0,  
    &&\text{if } t\! \in \! 
			\left\{N_{u},...,N_{p-1}\right\}  \\
\end{aligned}
\end{cases}
\label{eq:Var(h) dynamics cases}
\end{equation} 
In the above equations, $c^{(h)}_k$  represents the PCE coefficients of the nonlinear inequality constraints. 
The moments $\mathbb{E}[\Phi_{\alpha_k}(\boldsymbol{w})^2], \forall k \in\{1,L-1\}$ are precomputed offline and remain consistent for all subsequent use. In the case of assuming Gaussian random variables and hence, Hermite polynomial basis is taken, $\mathbb{E}[\Phi_{\alpha_k}(\boldsymbol{w})^2] = 1 , \forall k \in\{1,L-1\}$ (see Corollary  9 and 10 in \cite{rahman2017wiener}). The PCE coefficients can be computed as:
\begin{equation}
    \boldsymbol{c^{(h)}} = [c^{(h)}_0, c^{(h)}_1,\ldots, c^{(h)}_{L-1}]^T = A \cdot \boldsymbol{\Tilde{h}}(\boldsymbol{\Tilde{X}}, \boldsymbol{u})
\end{equation}
Where $\boldsymbol{\Tilde{h}}(\boldsymbol{\Tilde{X}}, \boldsymbol{u}) = [\Tilde{h}(\boldsymbol{\Tilde{x}}_0, \boldsymbol{u}), \ldots, \Tilde{h}(\boldsymbol{\Tilde{x}}_{n_s}, \boldsymbol{u}) ]^T$ is a vector containing the inequality constraint applied to the individual PC sampling points $\Tilde{\boldsymbol{x}}_i$. \\

\begin{equation*}
\begin{aligned}
&\textbf{Problem 2} && \textbf{Deterministic surrogate for}\\ 
&&&\textbf{Stochastic Nonlinear MPC}\\ 
\end{aligned}
\end{equation*}
\begin{equation}
\begin{aligned}
&  \underset{\boldsymbol{x}(.), \boldsymbol{u}(.)}{\min} & & 
\begin{aligned}
    \int^{T_p}_{\tau=0} & l( \mathbb{E}[\boldsymbol{x}(\tau)],\boldsymbol{u}(\tau)) \space  d\tau \\ 
 & + m(\mathbb{E}[\boldsymbol{x}(T_p)])
\end{aligned}
  \\ 
& \text{subject to} & & \boldsymbol{x}_{0} \leq \boldsymbol{x}(0) \leq \boldsymbol{x}_{0} \text {, }  \\
& & & \dot{\boldsymbol{x}}(t) = f(\mathbb{E}[\boldsymbol{x}(t)],\boldsymbol{u}(t)) \text {, } & t \in[0, T_p), \\
& & & \underline{\boldsymbol{g}} \leq g(\mathbb{E}[\boldsymbol{x}(t)], \boldsymbol{u}(t)) \leq \bar{\boldsymbol{g}},& t \in[0, T_p), \\
& & & 
\begin{aligned}
   \underline{\boldsymbol{h}} \leq & \mathbb{E}[h(\boldsymbol{x},\boldsymbol{u})] \\ & + \kappa\sqrt{\text{Var}[h(\boldsymbol{x},\boldsymbol{u})]}
\leq \bar{\boldsymbol{h}},
\end{aligned}
& t \in[0, T_p), \\
& & & \underline{\boldsymbol{x}} \leq J_{\mathrm{bx}} \space \mathbb{E}[\boldsymbol{x}(t)] \leq \bar{\boldsymbol{x}}, & t \in[0, T_p), \\
& & & \underline{\boldsymbol{u}} \leq J_{\mathrm{bu}} \space \boldsymbol{u}(t)\leq \bar{\boldsymbol{u}}, & t \in[0, T_p), \\
& & & \underline{\boldsymbol{g}}^{\mathrm{e}} \leq g^{\mathrm{e}}(\mathbb{E}[\boldsymbol{x}(T_p)]) \leq \bar{\boldsymbol{g}}^{\mathrm{e}} \\
& & & 
\begin{aligned}
    \underline{\boldsymbol{h}}^{\mathrm{e}} \leq &
\mathbb{E}[h^{\mathrm{e}} (\boldsymbol{x}(T_p))] \\ &+ \kappa\sqrt{\text{Var}[h^{\mathrm{e}} (\boldsymbol{x}(T_p))]}
\leq \bar{\boldsymbol{h}}^{\mathrm{e}}
\end{aligned}
\\
& & & \underline{\boldsymbol{x}}^{\mathrm{e}} \leq J_{\mathrm{bx}}^{\mathrm{e}} \space \mathbb{E}[\boldsymbol{x}(T_p)]\leq \bar{\boldsymbol{x}}^{\mathrm{e}} \\
\end{aligned}
\label{eq:SNMPC problem}
\end{equation}
In Problem 2, we present the formulation of the transformed SNMPC problem. It includes a terminal cost and terminal constraints. The cost function takes into account uncertainty through the expectations of the states, and only the initial condition incorporates the nominal state. Furthermore, the problem involves nonlinear system dynamics and addresses nonlinear constraints in two distinct ways. First, it deals with hard nonlinear constraints on the expectations of the states, denoted as $g$ and $g^{\mathrm{e}}$. Second, it handles nonlinear chance constraints through estimated deterministic surrogates in expectation and variance (\ref{eq:deterministic h}). Additionally, it gives the possibility to impose hard linear constraints on the states' expectations and on the control inputs.
Finally, algorithm \ref{alg:SNMPC} illustrates the online implementation of our SNMPC approach in a receding horizon fashion.
\begin{algorithm}[ht]
\caption{PCE-SNMPC with an Uncertainty Propagation Horizon}\label{alg:SNMPC}
\algrenewcommand\algorithmicrequire{\textbf{Init}}
\begin{algorithmic}[1]
\State Set NMPC parameters: $T_p$, $T_{s}$, $Q$, $Q_e$ and $R$
\State Set SNMPC specific parameters: $n_{s}$, $\boldsymbol{\sigma}_{w,\text{SNMPC}}$ and $T_u$
\For{$N$ online steps}
    \State Update SNMPC's initial state $\boldsymbol{x}_0$ with current measurements
    \State Sample $n_{s}$ collocation points $\boldsymbol{\tilde{W}}$ 
    \State Generate $\boldsymbol{\Tilde{X}}$ around the current initial state $\boldsymbol{x}_0$
    \State Compute the PCE matrix solution $A$ 
    \State Update current SNMPC reference
    \For{$j \leq N_p$}
        \State Propagate the states according to (\ref{eq:x dynamics cases})
        \State Estimate nonlinear constraints according to (\ref{eq:deterministic h})
    \EndFor
    \State Solve the SNMPC problem
    \State Apply the first control input $\boldsymbol{u}^*_0$ on the real system
\EndFor
\end{algorithmic}
\end{algorithm}


%% file: sections/traj_following.tex
We address the challenge of controlling the combined longitudinal and lateral motion of a full-scale autonomous vehicle to follow a reference trajectory. Our experimental platform is EDGAR, the TUM research vehicle \cite{karle2023edgar}, a Volkswagen T7 Multivan customized for the development of autonomous driving software. The system is subject to several types of disturbances and especially state estimation uncertainties $\boldsymbol{x}_{t+1} = f(\boldsymbol{x}_{t},\boldsymbol{u}) + \boldsymbol{w}_t$, where $\boldsymbol{w}_t$ denotes the disturbances vector.
\subsection{Cost Function}
We define the vehicle state vector as follows:
\begin{equation}
\begin{aligned}
\boldsymbol{x} &= [x_{\text{pos}},\space y_{\text{pos}},\space \psi,\space v_{\text{lon}},\space v_{\text{lat}},\space \dot{\psi},\space \delta_f,\space a]^T
\end{aligned}
\end{equation}
In this vector, $x_{\text{pos}}$ and $y_{\text{pos}}$ represent the x- and y-coordinates of the ego vehicle, $\psi$ is the yaw angle, $v_{\text{lon}}$ and $v_{\text{lat}}$ indicate velocities in the longitudinal and lateral directions, $\dot{\psi}$ represents the yaw rate, $\delta_f$ corresponds to the steering angle at the front wheel, and $a$ signifies acceleration. 
The control vector is defined as $u = [j, \space \omega_f ]^T$, where $j$ represents the longitudinal jerk, and $\omega_f$ represents the steering rate at the front wheel.

The stage cost is defined as a nonlinear least square function: $l(\boldsymbol{x}, \boldsymbol{u})=\frac{1}{2}\|\boldsymbol{y}(\boldsymbol{x},\boldsymbol{u})-\boldsymbol{y}_{\mathrm{ref}}\|_W^2$. Similarly, the terminal cost is formulated as $m(\boldsymbol{x})=\frac{1}{2}\|\boldsymbol{y}^e(\boldsymbol{x})-\boldsymbol{y}^e_{\mathrm{ref}}\|_{W^e}^2$. Here, $W$ and $W_e$ represent the weighting matrices for the stage and terminal costs, respectively. $W$ is computed as $W = \text{diag}(Q,R)$, where $Q$ and $R$ are matrices for states and inputs weighting, while $W_e$ is defined as $W_e = Q_e$. The cost terms are defined as follows: 
\begin{equation}
  \begin{aligned}
\boldsymbol{y}(\boldsymbol{x},\boldsymbol{u}) &= [x_{\text{pos}},\space y_{\text{pos}},\space \psi,\space v_{\text{lon}},\space  j, \space \omega_f] \\
\boldsymbol{y}_{\text{ref}} &= [x_{\text{pos,ref}},\space y_{\text{pos,ref}},\space \psi_{\text{ref}},\space v_{\text{ref}},\space  0,\space 0] \\
\boldsymbol{y}^e_{\text{ref}} &= [x_{\text{pos,ref}}^e,\space y_{\text{pos,ref}}^e,\space \psi^e_{\text{ref}},\space v^e_{\text{ref}}]\\
\end{aligned}  
\end{equation}

To determine appropriate matrices $Q$ and $Q_e$ and $R$, we employ Multi-Objective Bayesian Optimization:
$
    Q = Q_e =  \text{diag} ( 
        2.8, \space 
        2.8, \space 
        0.4, \space 
        0.2
    )
$
and 
$
    R= \text{diag} ( 
        38.1, \space 
        101.4
    )
$.
\subsection{Nonlinear Prediction Model}
We adopt a dynamic nonlinear single-track model combined with Pacejka Magic Formula \cite{pacejka1997magic} to account for essential dynamic effects. The system dynamics are defined as: $f(\boldsymbol{x},\boldsymbol{u}) =$
\begin{equation}
\begin{aligned}
\begin{bmatrix}
v_{lon} \cos(\psi) - v_{lat} \sin(\psi) \space \\
v_{lon} \sin(\psi) + v_{lat} \cos(\psi) \space \\
\dot{\psi} \\
\frac{1}{m}\left(F_{x_r} - F_{y_f} \sin(\delta_f) + F_{x_f}  \cos(\delta_f) + m  v_{lat} \dot{\psi}\right) \\
\frac{1}{m}\left(F_{y_r} + F_{y_f} \cos(\delta_f) + F_{x_f} \sin(\delta_f) - m v_{lon} \dot{\psi}\right) \\
\frac{1}{I_z}\left(l_f \left(F_{y_f} \cos(\delta_f) + F_{x_f} \sin(\delta_f)\right) - l_r  F_{y_r}\right) \\
\omega_f \\
j  
\end{bmatrix} 
\end{aligned}
\end{equation}
For the lateral forces $Fy_{\{f,r\}}$, we account for the combined slip of lateral and longitudinal dynamics following the approach presented in \cite{raji2022motion}, where ${{f,r}}$ refers to the front or rear tires.
\begin{equation}
\begin{aligned}
Fy_{\{f,r\}} &= F_{{\{f,r\},\text{tire}}} \cos\left(\arcsin\left(F_{x_{\{f,r\}}}/F_{max_{\{f,r\}}}\right)\right)
\end{aligned}
\end{equation}
To avoid singularity problems, we clip $F_{x_{\{f,r\}}}/F_{max_{\{f,r\}}}$ at 0.98 \cite{raji2022motion}. 
The lateral front and rear tire forces are defined using the reduced Pacejka magic formula \cite{pacejka1997magic}: 
\begin{equation}
\begin{aligned}
F_{{\{f,r\},\text{tire}}} &= D_{\{f,r\}} \sin(C_{\{f,r\}} \arctan(B_{\{f,r\}} \alpha_{\{f,r\}} \\
&- E_{\{f,r\}} (B_{\{f,r\}} \alpha_{\{f,r\}}- \arctan(B_{\{f,r\}} \alpha_{\{f,r\}}))))
\end{aligned}
\end{equation}
The side slip angles are defined as follows: 
\begin{equation}
\begin{aligned}
\alpha_f &= 
\delta_f - \arctan\left((v_{lat}+l_f \cdot \dot{\psi})/v_{lon}\right)  \\
\alpha_r &= 
\arctan\left((l_r \cdot \dot{\psi} - v_{lat}) /v_{lon}\right)
\end{aligned}
\end{equation}
The tire sideslip angle formula has a singularity issue with longitudinal velocity \cite{smith1995effects}. To address this, we assume that the tire sideslip angle is negligible at low velocities, avoiding the need for two different models.
The longitudinal forces are defined as following:
\begin{equation}
\begin{aligned} 
Fx_f &= - Fr_f \\
Fx_r &= F_d  - Fr_r  - F_{aero}\\
\end{aligned}
\end{equation}
Here, the driving force at the wheel is defined as $ F_d = m \cdot a$ and the rolling resistance forces \cite{gerdts2003single} are defined as
$Fr_{\{f,r\}} = fr \cdot F_{z,{\{f,r\}}}$. The rolling constant $f_r$ is defined as $fr = fr_0 + fr_1 \cdot \frac{v}{100} + fr_4 \cdot \left(\frac{v}{100}\right)^4$, where $v$ represents the absolute velocity in $\unit{\kilo\meter\per\hour}$ \cite{gerdts2003single}.
The aerodynamic force is calculated as $F_{aero} = 0.5\cdot \rho \cdot S \cdot Cd \cdot v_{lon}^2 $\cite{gerdts2003single} and $F_{z,\{f/r\}}$ represents the vertical static tire load at the front and rear axles $F_{z,{\{f,r\}}} = \frac{m \cdot g \cdot l_{\{r,f\}}}{l_f + l_r}$. 

To identify the prediction model parameters, we perform ISO 4138-compliant steady-state circular driving tests. We refer to Section V-I in \cite{karle2023edgar} for the parameter values of the nonlinear single track and Pacejka tire model used in this work.

\subsection{Constraints}
We formulate the combined longitudinal and lateral acceleration potential limits for the SNMPC (Problem 2) as a nonlinear probabilistic constraint using Eq.\ref{eq:deceleration constraints} and for the nominal NMPC (Problem 1) as a nonlinear hard constraint: 
\begin{equation}
    h(\boldsymbol{x}, \boldsymbol{u}) = (a_{lon}/a_{x_{max}})^2 + (a_{lat}/a_{y_{max}})^2
    \label{eq: acc potential}
\end{equation}
Here, the longitudinal acceleration is $a_{lon} = a$, and the lateral acceleration is $a_{lat} = v_{lon} \dot{\psi}$. The upper and lower bounds for $h$ are $\bar{h} = 1$ and $\underline{h} = 0$. We adapt the maximum allowed values based on the limits defined by the vehicle's actuator interface software. Specifically, we set $a_{y_{max}} = 5.866 \unit{\meter\per\second\squared}$, while $a_{x_{max}}$ varies based on the current velocity. When decelerating, $a_{x_{max}}$ is defined as:
\begin{equation}
\begin{aligned}
a_{x_{max}} = 
\begin{cases}
     |-4.5\unit{\meter\per\second\squared}|,  &\text{if } 0\leq v_{lon} \leq 11 \unit{\meter\per\second} \\
     |-3.5\unit{\meter\per\second\squared}|,  &\text{if } 11 \unit{\meter\per\second} < v_{lon} \leq 37.5 \unit{\meter\per\second}
\end{cases}
\end{aligned}
\label{eq:deceleration constraints}
\end{equation}
And during acceleration:
\begin{equation}
\begin{aligned}
a_{x_{max}} = 
\begin{cases}
    3 \unit{\meter\per\second\squared}, & \text{if } 0\leq v_{lon} \leq 11 \unit{\meter\per\second} \\
     2.5\unit{\meter\per\second\squared}, & \text{if } 11 \unit{\meter\per\second} < v_{lon} \leq 37.5 \unit{\meter\per\second}
\end{cases}
\end{aligned}
\label{eq:acceleration constraints}
\end{equation}
Additionally, we impose linear hard constraints on the steering angle and steering rate at the front wheel:

\begin{equation}
\begin{aligned}
-0.61 \unit{\radian}\leq & \mathbb{E}[\delta_f] \leq 0.61\unit{\radian} \\
-0.322\unit{\radian\per\second}\leq &\omega_f \leq 0.322\unit{\radian\per\second}
\end{aligned}
\label{eq:hard constraints}
\end{equation}
\subsection{SNMPC configuration}

For both our nominal NMPC and SNMPC, we have set the following parameters: the sampling time $T_s = 0.08 \unit{\second}$, the prediction horizon $T_p = 3.04 \unit{\second}$. Specifically for SNMPC, we have configured the uncertainty propagation horizon $T_u = 0.4 \unit{\second}$ and the desired probability of constraint violation as $p = 0.8$. We operate under the assumption that the system states, namely $v_{lon}$, $v_{lat}$, and $\dot{\psi}$, are subject to Gaussian disturbances, as our experiments have demonstrated that this assumption leads to the most substantial improvements in closed-loop performance and computational load. In the case of SNMPC, we assume the following standard deviations:
\begin{equation}
\begin{aligned}
    \boldsymbol{\sigma}_{w}^{\text{SNMPC}} &= [\sigma_{\text{vlon}}, \sigma_{\text{vlat}}, \sigma_{\dot{\psi}}] ^T \\ &= [0.8\unit{\meter\per\second}, 0.35\unit{\meter\per\second}, 0.035\unit{\radian\per\second}]^T
\end{aligned}
\end{equation} 
Thus, the total number of uncertain system parameters is $n_w = 3$, and we employ Hermite polynomials. We have chosen an expansion degree of $d_{PC,max} = 2$ with a sample count of $n_s = 10$. Consequently, the number of terms in the truncated PC expansion equals $L = 10$. Importantly, when deployed on the vehicle, the state estimation module determines the standard deviations based on sensor input.

%% file: sections/sim_results.tex
In this section, we conduct a performance comparison between our SNMPC approach and an equivalent nominal NMPC, both subjected to significant additive Gaussian disturbances affecting the state estimates. 
\subsection{Simulation Setup}
\label{subsec:simulation_setup}
We conducted our experiments on a standard laptop featuring an Intel i7-11850H 2.50GHz CPU and 16GB of RAM. We implemented both NMPC and SNMPC using the ACADOS library \cite{Verschueren2021} in Python 3.9, which automatically generates C-Code for efficient execution. Our prediction- and simulation models are formulated with CasADi \cite{Andersson2019}. For both MPCs, we use the SQP RTI as an NLP Solver and HPIPM QP Solver with 50 as the number of maximum allowed iterations.

\begin{figure}[h]
\includegraphics[width=.48\textwidth]{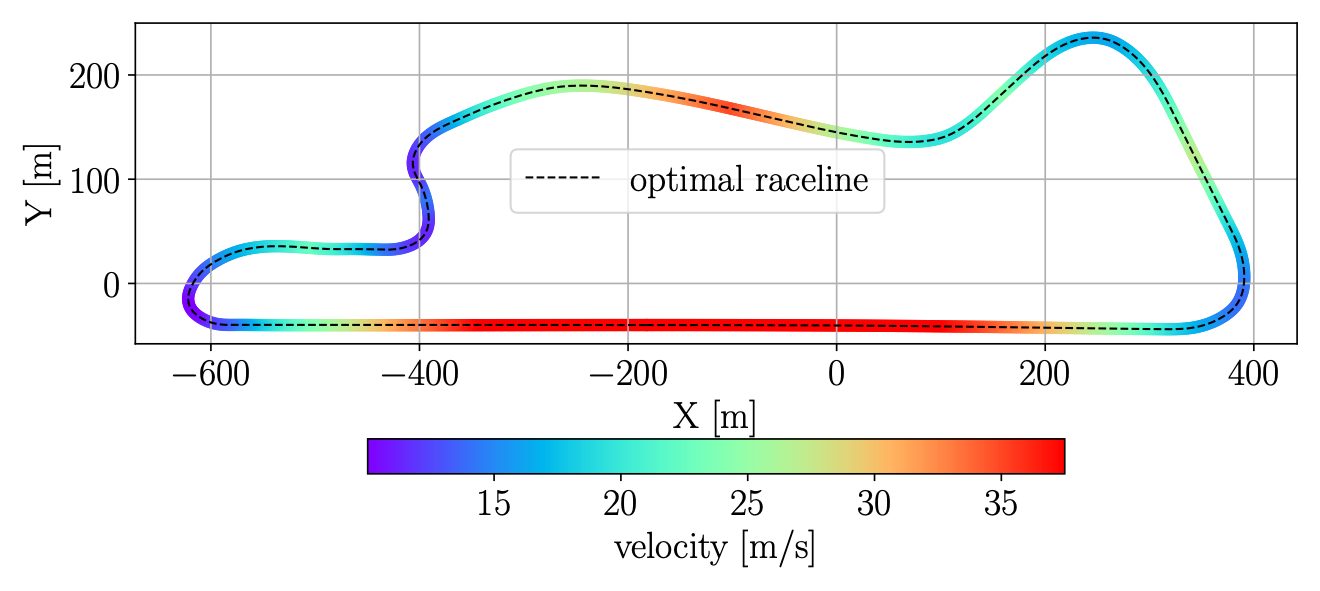}
\caption{Track layout and velocity profile at Monteblanco}
\label{fig:racetrack}
\end{figure}

We generate the SNMPC reference trajectory for a real-world racetrack called Monteblanco based on minimum curvature optimization from the global race trajectory optimization framework \cite{heilmeier2019minimum}. This framework allows us to generate the optimal trajectory using the full dynamics according to the given limits. This allows us to analyze the performance of the controller near its constraints. We take into account the vehicle interface limits defined in (\ref{eq: acc potential}-\ref{eq:hard constraints}). Our TUM research vehicle interface allows a maximum velocity of $37.5\unit{\meter\per\second}$. Figure \ref{fig:racetrack} depicts the layout for Monteblanco, the optimal raceline and the reference velocity profile for the TUM research vehicle.\\
To simulate disturbances, we add Gaussian noise to the measured states with the following standard deviations:
\begin{equation}
    \begin{aligned}
    \boldsymbol{\sigma}_{w}^{\text{sim}}=& [\sigma_x,\sigma_y,\sigma_{\psi},\sigma_{vlon},\sigma_{vlat},\sigma_{\dot{\psi}},\sigma_{\delta_f}]^T \\ =&[0.1\unit{\meter},0.1\unit{\meter},0.05\unit{\radian},0.8\unit{\meter\per\second}, \\ & 0.35\unit{\meter\per\second},0.035\unit{\radian\per\second},0.01\unit{\radian}]^T
    \end{aligned}
\label{eq:sim_dist_normal_setup}
\end{equation}
To ensure a fair comparison, both SNMPC and NMPC are subjected to the same disturbance realization, i.e., $\boldsymbol{w}_t^{\text{SNMPC}} = \boldsymbol{w}_t^{\text{NMPC}}, \space \forall t \in \{0,\ldots,N\}$.
 Additionally, we employed a simple moving average filter to smooth the input signals, using the following window sizes for each input state: $[1,1,4,2,2,3,4,2]^T$.

In section \ref{subsec:traj_foll_performance}, we compare the performance of the proposed SNMPC with the nominal NMPC both with and without disturbance influence by simulating the vehicle following the optimal raceline for $120\unit{\second}$. Notably, in these benchmarks, we maintained identical configurations and parameters for both NMPC and SNMPC. We also demonstrate the importance of the proposed UPH approach in Section \ref{subsec:impact_uph}. 
\subsection{Trajectory Following Performance}
\label{subsec:traj_foll_performance}
\begin{figure}[ht]
\includegraphics[width=.48\textwidth]{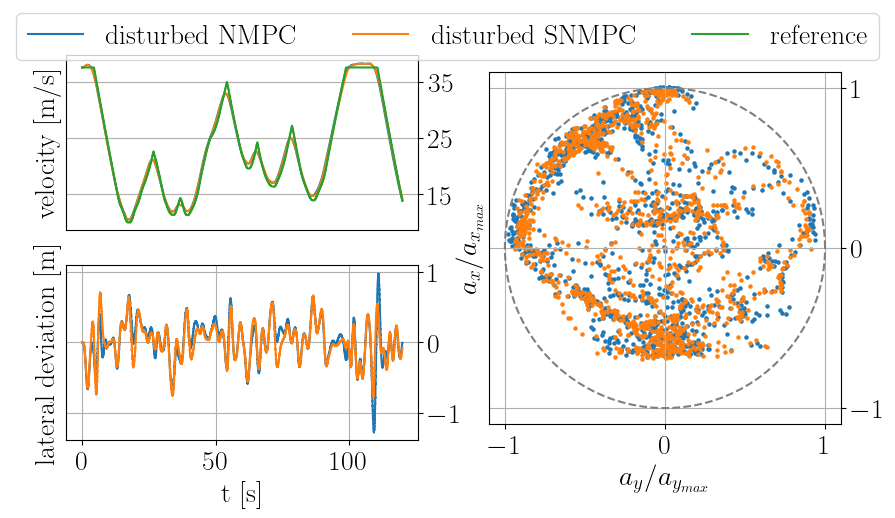}
\caption{MPC's closed-loop performance subject to the disturbance in Monteblanco racetrack: velocity, lateral deviation and gg-diagram.}
\label{fig:benchmark_disturbed_nmpcs_gg}
\end{figure}

\begin{figure}[h]
\includegraphics[width=.48\textwidth]{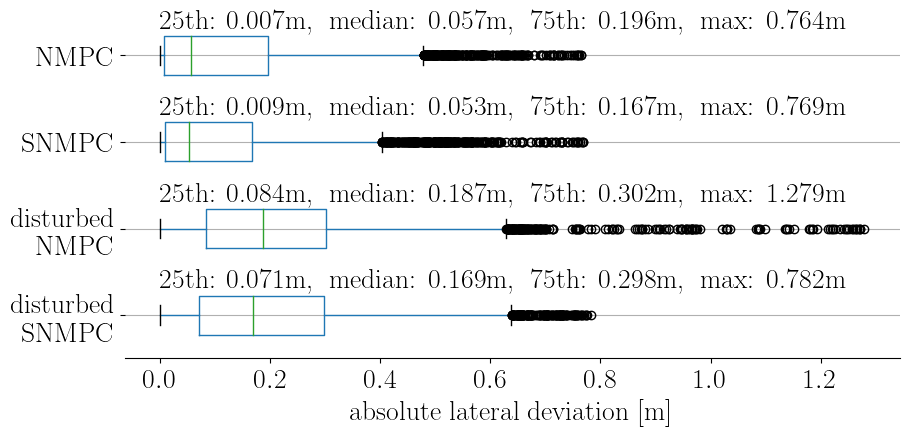}
\caption{Effect of the disturbance on the NMPC and SNMPC w.r.t. the absolute lateral deviation.}
\label{fig:benchmark4nmpcs}
\end{figure}

Figure \ref{fig:benchmark_disturbed_nmpcs_gg} provides a comparative analysis of the Nominal NMPC and SNMPC under the influence of disturbances as described in (\ref{eq:sim_dist_normal_setup}). \\
The acceleration plot (gg-diagramm) shows that the SNMPC operates within the system limits significantly more effectively than the Nominal NMPC. The latter frequently violates system constraints, as observed in the upper center of the gg-diagram, while the SNMPC, employing estimated nonlinear chance constraints as defined in (\ref{eq:deterministic h}) and (\ref{eq: acc potential}), exhibits no such violations. This performance disparity underscores the SNMPC's ability to effectively account for uncertainties and propagate them. \\
The velocity plot reveals that both controllers closely track the reference velocity, displaying similar behavior. However, a big performance contrast becomes apparent in the lateral deviation plot. \\
Figure \ref{fig:benchmark4nmpcs} emphasizes the lateral behavior of the two controllers. The benchmarking is conducted under two scenarios: one without disturbance and one with disturbance. 
When comparing the controllers under normal operating conditions (without disturbance), the SNMPC exhibits a slightly better performance.\\
However, when disturbances are introduced into the system, we observe a notable difference in controller performance. The nominal NMPC's performance degrades significantly. In contrast, the SNMPC demonstrates remarkable resilience to disturbances, maintaining a consistently lower 25th, median, and 75th percentiles of lateral deviation. This substantial advantage is further emphasized when we consider the maximum deviation. \\
The absolute maximum lateral deviation represents the worst-case scenario in terms of how far the vehicle can deviate from its intended trajectory. Minimizing this deviation is crucial to ensure that the vehicle stays within its lane and avoids collisions with other vehicles, obstacles, or pedestrians.\\
While the Nominal NMPC reaches a peak deviation of 1.279 meters, the SNMPC achieves a markedly lower 0.782 meters. This represents an approximately \textbf{61\%} improvement in limiting the maximum lateral deviation, underscoring the SNMPC's superior performance in challenging scenarios. \\
Remarkably, even in the presence of significant disturbances, the SNMPC not only significantly outperforms the NMPC but also maintains a performance close to its nominal conditions, with only a marginal degradation of \textbf{1.6\%} in terms of maximum deviation, indicating that it is not overly influenced by outliers or extreme events. 
\begin{table}[h]
\centering
\caption{Solver computational time}
\label{tab:cpu_time}
\begin{tabular}{@{}lllll@{}}
\toprule
           & NMPC & SNMPC  & \\ \midrule
Maximum {[}ms{]} & 2.147  & 10.236    \\
Mean {[}ms{]}    & 1.049  & 5.8   &         \\ \bottomrule
\end{tabular}
\end{table}
The computational times, as summarized in Table \ref{tab:cpu_time}, provide valuable insights into the efficiency and responsiveness of both controllers. The NMPC demonstrates remarkable computational efficiency, capable of solving problems at a minimum frequency of $456 \unit{\hertz}$. Given its stochastic modeling complexities, The SNMPC has a relatively higher computational load, achieving a minimum problem-solving frequency of $97 \unit{\hertz}$.
Despite the increased computational demands of the SNMPC, our proposed approach still aligns perfectly with our real-time control requirements. Specifically, it meets the critical update frequency of $50 \unit{\hertz}$ demanded by the vehicle control interface. This underscores the practical viability of our approach in the context of the trajectory following use case.

\subsection{Impact of the Uncertainty Propagation Horizon}
\label{subsec:impact_uph}
In this section, we highlight the pivotal role of constraining the Uncertainty Propagation Horizon (UPH) in enhancing the feasibility of control problems subjected to disturbances.
We conduct two experiments within the context of our simulation setup (Section \ref{subsec:simulation_setup}), both employing the SNMPC. These experiments compare the problem feasibility under two distinct UPH settings: the first experiment assumes a UPH equal to the entire prediction horizon (UPH = $T_p$), while the second employs a shorter UPH of 0.8 seconds.

\begin{figure}[ht]
\includegraphics[width=.48\textwidth]{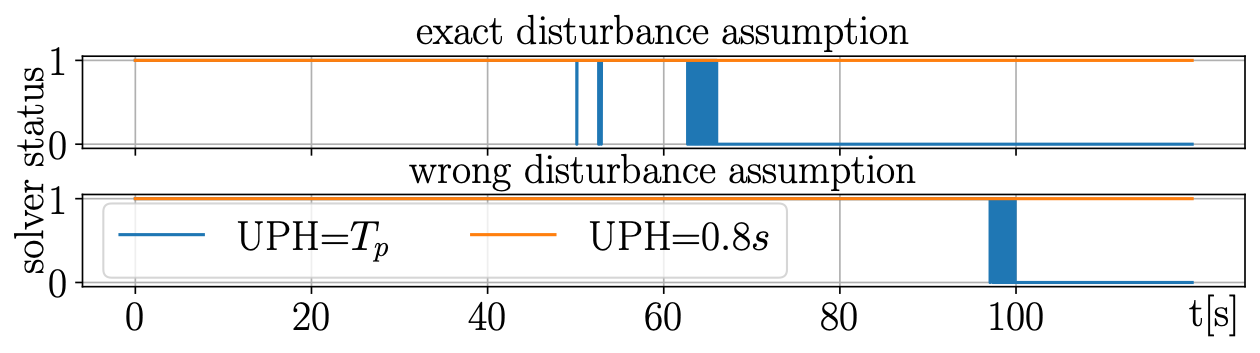}
\caption{Status of the solver for a short- and a long UPH subject to \underline{exact} (upper) and \underline{wrong} (lower) disturbance assumption of the SNMPC.}
\label{fig:solver_status}
\end{figure}

In the first experiment, we introduce significant additive Gaussian disturbances and configure the SNMPC with precise disturbance assumptions that match the simulation model. Specifically, the disturbance standard deviations are set as follows:
\begin{equation}
    \begin{aligned}
    \boldsymbol{\sigma}_{w}^{\text{sim}}=& [0.3\unit{\meter},0.3\unit{\meter},0.05\unit{\radian},0.8\unit{\meter\per\second}, \\ & 0.8\unit{\meter\per\second},0.1\unit{\radian\per\second},0.01\unit{\radian}]^T\\
    \boldsymbol{\sigma}_{w}^{\text{SNMPC}} =& [0.8\unit{\meter\per\second}, 0.8\unit{\meter\per\second}, 0.1\unit{\radian\per\second}]^T
    \end{aligned}
    \label{eq:config_right_w_assumption}
\end{equation}
In Figure \ref{fig:solver_status}, we present the solver's status throughout the simulation, where '1' indicates that the HPIPM solver successfully found a solution, and '0' denotes when a solution could not be found. Notably, propagating uncertainties through the entire prediction horizon ($T_p$) leads to an infeasible problem. However, restricting uncertainty propagation to a shorter UPH ensures problem feasibility at each time step.\\
In the second experiment, we modify the disturbance configuration as per (\ref{eq:config_wrong_w_assumption}), introducing incorrect SNMPC disturbance assumptions:
\begin{equation}
  \begin{aligned}
  \boldsymbol{\sigma}_{w}^{\text{sim}}=& [0.1\unit{\meter},0.1\unit{\meter},0.05\unit{\radian},0.8\unit{\meter\per\second}, \\ & 0.35\unit{\meter\per\second},0.035\unit{\radian\per\second},0.01\unit{\radian}]^T\\
    \boldsymbol{\sigma}_{w}^{\text{SNMPC}} =& [0.8\unit{\meter\per\second}, 0.8\unit{\meter\per\second}, 0.1\unit{\radian\per\second}]^T
\end{aligned}
    \label{eq:config_wrong_w_assumption}
\end{equation}
Figure \ref{fig:solver_status} illustrates how a shorter UPH renders the control problem feasible even in the presence of incorrect SNMPC disturbance assumptions.




